\documentclass[aps,prx,amsmath,twocolumn,groupedaddress,letterpaper,floatfix]{revtex4-2}
\usepackage{graphicx,color}
\usepackage{verbatim}
\usepackage{amssymb}   
\usepackage{amsmath}
\usepackage{amsfonts}
\usepackage{mathdots}
\usepackage{hyperref}
\usepackage{gensymb}
\usepackage{epsfig}
\usepackage{hyperref}
\definecolor{myblue}{RGB}{46, 48,146}
\hypersetup{colorlinks=true,linkcolor=myblue,citecolor=myblue,urlcolor=myblue, linktocpage}

\newcommand{\YM}[1]{\textcolor{black}{#1}} 
 
\newcommand{\CP}[1]{\textcolor{black}{#1}}
\usepackage{braket}
\usepackage{bm}
\begin{document}
	\title{Topological $p_x+ip_y$ inter-valley coherent state in Moir\'e MoTe$_2$/WSe$_2$ heterobilayers}

	\author{Ying-Ming Xie$^{1,2}$}
	\thanks{These authors contributed equally to this work.} 
	\author{Cheng-Ping Zhang$^1$}
	\thanks{These authors contributed equally to this work.} 
	\author{K. T. Law$^1$} \thanks{Corresponding author.\\phlaw@ust.hk}
	
	\affiliation{$^1$Department of Physics, Hong Kong University of Science and Technology, Clear Water Bay, Hong Kong, China}
	\affiliation{$^2$RIKEN Center for Emergent Matter Science (CEMS), Wako, Saitama 351-0198, Japan} 	
	
	\date{\today}
	\begin{abstract}
		Recently, a quantum anomalous Hall (QAH) state was observed in AB stacked moir\'e MoTe$_2$/WSe$_2$ heterobilayers at half-filling.  More recent layer-resolved magnetic circular dichroism (MCD) measurements revealed that spin-polarized moir\'e bands from both the MoTe$_2$ and the WSe$_2$ layers are involved at the formation of the QAH state. This scenario is not expected by existing theories. In this work, we suggest that the observed QAH state is a new state of matter, namely, a topological $p_x+ip_y$ inter-valley coherent state (TIVC). We point out that the massive Dirac spectrum of the MoTe$_2$ moir\'e bands, together with the Hund's interaction and the Coulomb interactions give rise to this novel QAH state. Through a self-consistent Hartree-Fock analysis, we study how this $p_x+ip_y$-pairing phase could be stabilized. Besides explaining several key features of the experiments, our theory predicts that the order parameter would involve the pairing of electrons and holes with a definite momentum mismatch such that the pairing would generate a new unit cell which is three times the size of the original moir\'e unit cell, due to the order parameter modulations. \end{abstract}

	\maketitle
	\section{\bf{Introduction}}
	The concepts of band topology and interaction effects have played an important role in the study and discovery of new phases of matter in physics in the past decades. Recently, moir\'e materials have become a fertile ground for studying the interplay between band topology and interactions \cite{Andrei2021}. For example, there have been experimental observations of quantum anomalous Hall (QAH)  states in twisted bilayer graphene \cite{David2019,Young2020} and AB stacked MoTe$_2$/WSe$_2$ heterobilayers \cite{Fai_nature2021,Andrea_2022}.  Unlike other QAH state platforms, such as magnetic doped topological insulator thin film \cite{YuRui2010,Cuizu_2013,Kang2015,Chang2015,Yasuda2020,Cuizu_2022} and MnBi$_2$Te$_4$ \cite{Yuanbo2020} in which the topological states can be understood from single-particle band structures, the bands in moir\'e materials are typically very narrow (several to tens of meV) so that Coulomb interactions cannot be neglected. A  possible path to obtain QAH states is to first create moir\'e Chern bands at each valley and lift the valley degeneracy through interactions. This would result in an interaction-driven valley-polarized QAH state. This idea has been proposed to explain the origin of the QAH states in graphene moiré superlattice \cite{PoHoiChun2018,Yahui2019,Senthil2019,Senthil2020_2,Zaletel2020,Chen2020,Macdonald2020}. However, it becomes highly controversial when applied to understand the recently discovered QAH states in MoTe$_2$/WSe$_2$ heterobilayes at half-filling $\nu=1$ (one hole per moir\'e unit cell) \cite{Fai_nature2021}.
	
	The discovery of QAH states in MoTe$_2$/WSe$_2$ heterobilayers is particularly surprising. This is due to the fact that there is a large energy misalignment between the valence band tops  of MoTe$_2$ and WSe$_2$, while the spin degeneracy is largely lifted by the giant Ising spin-orbit coupling \cite{XiaoDi2012,Lu2015,Xi2016}. The band alignment of the top moir\'e bands of MoTe$_2$ and WSe$_2$ layers is schematically depicted in Fig.~\ref{fig:fig1}a. As a result,  a transition metal dichalcogenide (TMD) heterobilayer is usually described by a single-band Hubbard model with valley degrees of freedom \cite{fengcheng2018,Yangzhang2020,Fai_hubbard2020, LiTingxin2021, Jin_Fai2021, Davydova2022}.   The moir\'e bands are thus expected to be topologically trivial. Immediately after the discovery of the QAH in MoTe$_2$/WSe$_2$ heterobilayers \cite{Fai_nature2021}, it was pointed out that lattice relaxation induced pseudomagnetic fields can make the top moir\'e bands topological \cite{Law_PRL2022}. At the same time, Zhang et. al. proposed an alternative way whereby the hybridization of the top moir\'e bands of MoTe$_2$ and WSe$_2$ can be achieved through displacement fields \cite{Fu_zhang2021}, resulting in Chern bands similar to the case of homobilayer TMDs \cite{fengcheng2019}. Subsequently, a series of works \cite{Devakul2021,Suying2021,DasSarma2021,Rademaker2021,Chang2022,Yahui2023} have been devoted to understanding the topological origin of AB-stacked MoTe$_2$/WSe$_2$ heterobilayers. However, the underlying mechanism still remains elusive.
	
	
	Interestingly, the very recent experiment~\cite{Faitalk} reported in MoTe$_2$/WSe$_2$ heterobilayers further sheds light on understanding the QAH state of this system. 
	Specifically, the layer-resolved magnetic circular dichroism (MCD) measurements clearly indicate that the QAH state at half-filling would appear when the top moir\'e  band of WSe$_2$ is hole-doped and both the top moir\'e  bands of MoTe$_2$ and WSe$_2$ layers would have the same spin ~\cite{Faitalk}. In other words, the MCD measurements suggest a band alignment scenario as depicted in Fig.~\ref{fig:fig1}b. As such, the QAH state involves the coherent superposition of electrons and holes from both valleys and layers with the same spin, which is unexpected from all previous theoretical proposals \cite{Law_PRL2022,Fu_zhang2021,Devakul2021,Suying2021,DasSarma2021,Rademaker2021,Chang2022}. Therefore, the discovery of this intriguing, interaction-driven, QAH state calls for a theoretical understanding of its microscopic origin. 
	
	
	In this work, we propose that the observed QAH state is a topological $p_x+ip_y$ inter-valley coherent (TIVC) state.  We first show that a Hund interaction would give rise to band alignment as depicted in Fig.~\ref{fig:fig1}b. The corresponding band structure from a continuum model with Hund's interaction is shown in Fig.~\ref{fig:fig1}d. Near half-filling, the state near the Fermi energy came from an electron pocket from the MoTe$_2$ layer and a hole pocket from the WSe$_2$ layer as depicted in Figs.~\ref{fig:fig1}d and \ref{fig:fig1}e. Importantly, by pairing electrons and holes with a momentum difference of $\bm{Q}$ as defined in Fig.~\ref{fig:fig1}e, Coulomb interactions would induce an order parameter which gaps out the electron and the hole bands as shown in Fig.~\ref{fig:fig2}. Moreover, the order parameter has a $p_x+ip_y$ momentum dependence which is inherited from the massive Dirac Hamiltonian of the MoTe$_2$ layer near the Fermi energy. This is a key finding of our theory. Finally, we calculate the phase diagram as a function of the interaction strength and the displacement field numerically and show that there is a large phase space in which the TIVC is possible to be stabilized.  Our calculations clearly show that the TIVC state can be induced by interactions when the WSe$_2$ bands are hole-doped, as observed experimentally.  Our theory thus identifies a microscopic route to generate the  TIVC state in moir\'e materials and explains several key features seen in recent experiments involving moir\'e MoTe$_2$/WSe$_2$ heterobilayers \cite{Fai_nature2021, Faitalk}. Importantly, we predict that the hybridization of the states with a momentum difference $\bm{Q}$ folds the bands and creates a new unit cell which is three times larger than the original moir\'e unit cell.
	
	\section{Continuum model}
	
\begin{figure}
		\centering
		\includegraphics[width=1\linewidth]{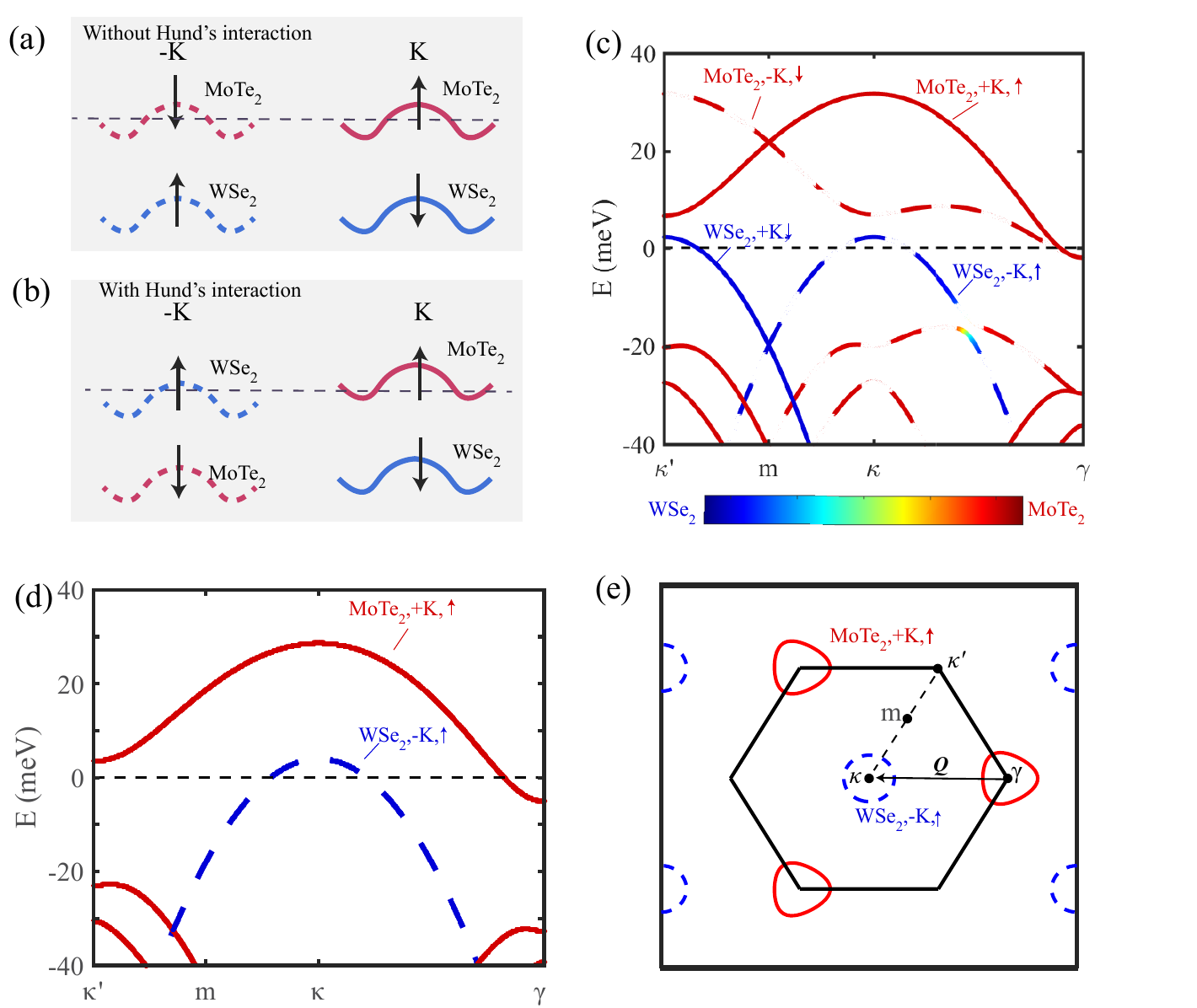}
		\caption{(a) A schematic plot of the alignment of the top moir\'e bands of different spin at two valleys before and after including the Hund's interaction. (b) The calculated moir\'e band from the continuum model, where the layer, spin, and valley information of these moir\'e bands are highlighted in the figure.  The red (blue) color labels the MoTe$_2$ (WSe$_2$) layer, respectively, while the rainbow color is due to the hybridization of MoTe$_2$ and WSe$_2$ layers induced by interlayer hopings. (c) The moir\'e bands with the spin degeneracy lifted by magnetization energy $M_z=50$ meV. (d) The Fermi contours of (c) at half-filling (dashed line).   Here the band offset energy $\delta E_D$ is taken as 25 meV for (c) and 20 meV for (d), respectively.  Other adopted parameters in this figure are  $m_b=0.65 m_0$ ($m_0$ is the electron mass), $m_t=0.35 m_0$ , $\phi_b=14^{\circ}$, $\lambda_b\kappa^3=5$, $V_b=10$ meV, $\lambda_t=0$, $V_t=0$,  $W=1.3$ meV. }
		\label{fig:fig1}
	\end{figure}

	We first present a continuum model which can properly describe the moir\'e bands of  AB-stacked heterobilayer MoTe$_2$/WSe$_2$ \cite{Fu_zhang2021,Fai_nature2021}.  The continuum model reads
	\begin{align}
		\mathcal{H}_{\tau}(\bm{r})=
		\begin{pmatrix}
			\mathcal{H}^{b}_{\tau}(\bm{r})+\frac{1}{2}\tau M_{z,b}
			& W_{\tau}(\bm{r})\\
			W_{\tau}^{\dagger}(\bm{r})
			& \mathcal{H}^{t}_{\tau}(\bm{r}) -\frac{1}{2}\tau M_{z,t}- \delta E_{D}
		\end{pmatrix}.\label{continuum_model}
	\end{align}
	Here,  the intralayer moir\'e Hamiltnoian near $\tau$-valley is given by $	\mathcal{H}^{l}_{\tau}=-\frac{\hat{\bm{p}}^2}{2m_{l}}+\tau\lambda_{l}(\hat{p}_x^3-3\hat{p}_x\hat{p}_y^2)+V_{l}(\bm{r})$, where $\bm{\hat{p}}=-i\hbar\bm{\nabla}$ is the momentum operator. The layer index $l=b/t$ denotes the MoTe$_2$/WSe$_2$ layer respectively and $m_{l}$ is the valence band effective mass of the $l$-th layer. The warping term that breaks the degeneracy of $\pm \kappa$ valley of the moir\'e Brillouin zone is present as indicated by the density functional theory (DFT) calculations \cite{Fu_zhang2021}. The intralayer moir\'e potential is
	$V_{l}(\bm{r})=2 V_{l}\sum_{j=1,3,5}\cos(\bm{G}_j\cdot\bm{r}+\phi_{l})$, while  the interlayer hopping is $W_{\tau}(\bm{r})=W(1+\omega^{\tau}e^{-i \bm{G}_2\cdot\bm{r}}+\omega^{2 \tau}e^{-i \bm{G}_3\cdot\bm{r}})$ with $\omega=e^{i2\pi/3}$, and the moir\'e reciprocal lattice vectors are $\bm{G}_{j}=\frac{4\pi}{\sqrt{3}a_{M}}[-\sin \frac{(j-1)\pi}{3},\cos \frac{(j-1)\pi}{3}]$ with $a_M\approx 5$ nm. The energy difference caused by the displacement field is denoted by $\delta E_{D}$.  With appropriate parameters (see Appendix A), this model captures the key features of the moir\'e bands from the DFT calculations without interaction effects \cite{Fu_zhang2021}. The band structure without the Hund's interaction ($M_z =0$) is schematically shown in Fig.~\ref{fig:fig1}a. This band structure is expected from the strong Ising SOC, time-reversal symmetry, and together with the AB-stacking of the TMD heterobilayer. For a clearer illustration,  the calculated  moir\'e bands from the continuum model (Eq.~\ref{continuum_model}) are shown in Fig.~\ref{fig:fig1}c. Here, the red and blue colors indicate states from the  MoTe$_2$ and WSe$_2$ layers respectively. The solid and dashed bands originate from the $ K $ and $-K$ valleys, respectively. Note that the separation between the top moir\'e bands of the WSe$_2$ and  MoTe$_2$ layers can be flexibly tuned by the displacement field. Without gating, the top moir\'e bands from the WSe$_2$ layer are expected to be about 200 $\sim$ 300 meV below the top moir\'e bands of the MoTe$_2$ layer \cite{Fai_nature2021,LiTingxin2021}.
	
	However, the band structure in Figs.~\ref{fig:fig1}a and \ref{fig:fig1}c are not consistent with the experimental findings which showed that at half--filling, the top moir\'e bands of the MoTe$_2$ and the WSe$_2$ layers have the same spin. Moreover, the QAH state appears when the top WSe$_2$ band becomes hole-doped. The new layer-resolved MCD measurements suggest a band alignment as shown in Fig.~\ref{fig:fig1}b. To take into account the spin polarization observed in the experiment ~\cite{Faitalk},  naively,  one can introduce opposite magnetization energy at two layers: $M_{z,b}$ and $M_{z,t}$ in the continuum model. As an illustration,  the spin degeneracy of the $K$ and $-K$ valleys lifted by  $M_{z,1}=M_{z,2}=50$ meV,  is    plotted in Fig.~\ref{fig:fig1}d. The corresponding Fermi pockets at half-filling are depicted in Fig.~\ref{fig:fig1}e, which include a hole pocket near $\kappa$ (in blue) arising from moir\'e band maximum of WSe$_2$ layer and three equivalent electron pockets near $\gamma$ (in red) arising from the  moir\'e band minimum of MoTe$_2$ layer with the same spin. Such a scenario is compatible with the MCD observation. \YM{It is worth noting that the top moir\'e band in Fig.~\ref{fig:fig1}d  is still topological trivial in this case before the formation of TIVC states. }
	\begin{figure}
		\centering
		\includegraphics[width=1\linewidth]{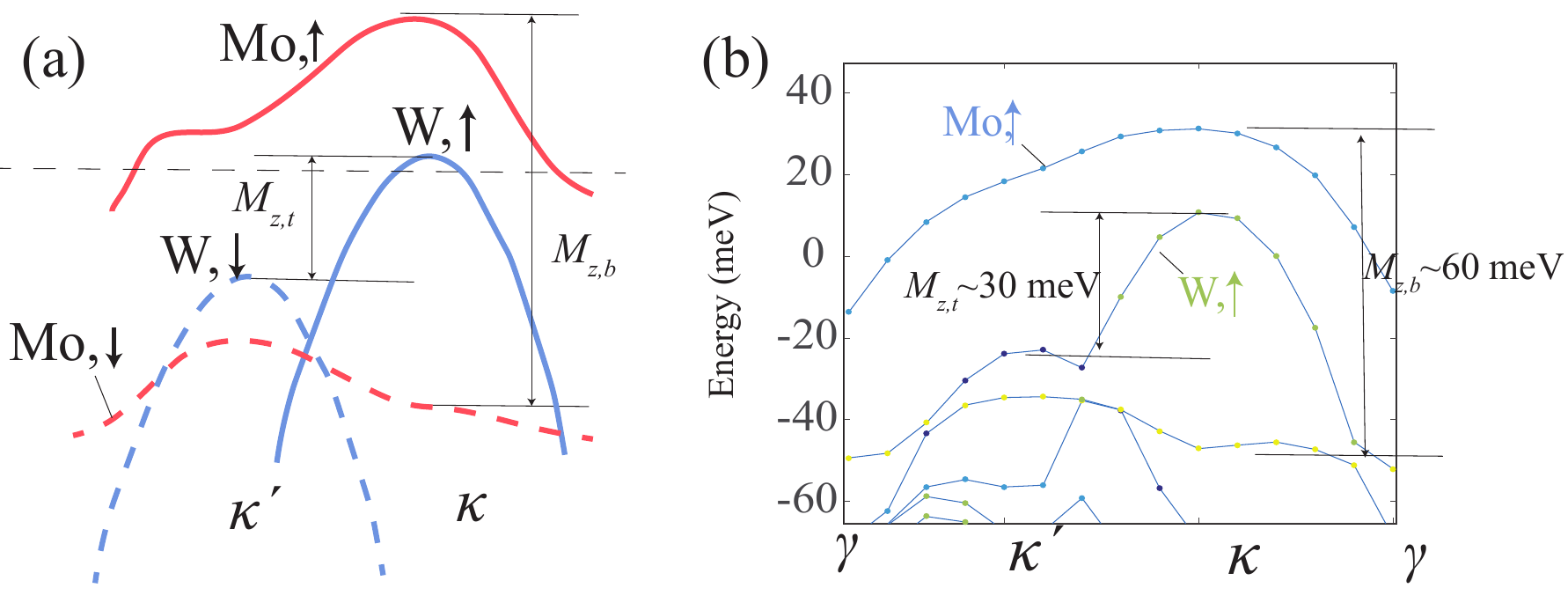}
		\caption{ \YM{The Hund's splitting induced by valley polarization effects. (a) schematically shows how the moir\'e bands behave in the presence of opposite valley polarization at two layers, which mimics the band alignment given in (b). (b)  shows the resulting moire bands after including the  Fock energy in each valley self-consistently with a Coulomb interaction set by a dielectric constant 5, screening length 5 nm. } }
		\label{fig:fig5}
	\end{figure}

	
	\YM{Next, we propose a  possible physical origin for this sizable  Hund's splitting (tens of meV).  We observe that the main effects of the magnetization $M_{z,b}$ and $M_{z,t}$ are to introduce valley polarization on the MoTe$_2$ layer and WSe$_2$ layer respectively. The valley polarization can be easily induced by the intra-valley Fock energy term under strong Coulomb interaction, which spontaneously breaks time-reversal symmetry by lifting the valley degeneracy \cite{Lee2019,Law_PRL2022}. Due to the spin-valley locking behavior in AB stacked heterobilayer, a sizable  opposite valley polarization at two layers can drive spin-polarized moir\'e bands near the Fermi energy (see Fig.~\ref{fig:fig5}a).}
	
	\YM{To justify this argument, the Fock energy term in each valley that gives rise to the valley polarization is evaluated in the plane wave basis (for details see Appendix B). Fig.~\ref{fig:fig5}b shows the moire bands after including the Fock energy using a reasonable Coulomb interaction. In this case,  it can be seen that  $M_{z,b}\sim 60$  meV, $M_{z,t}\sim 30$  meV.  $M_{z,b}$  is larger than $M_{z,t}$ because the Mo bands have narrower bandwidth.  The band alignment of  Fig.~\ref{fig:fig5}b  can be compared to the schematic plot shown in Fig.~\ref{fig:fig5}a. The bands near Fermi energy after valley polarization are consistent with what we consider in Fig.~\ref{fig:fig1}b and 1d. Note that in our theory,  the exact number of $M_{z,b}$ and $M_{z,t}$ are not that essential as long as they can create the spin-polarized bands near Fermi energy. }

	\section{Low energy effective Hamiltonian for spin-polarized moir\'e bands.} The next question is, starting from the band structure of Figs.~\ref{fig:fig1}b and \ref{fig:fig1}d, how can a topological QAH state arise from interactions? This is the question which we are mainly trying to address in this work. In this section, we develop a $\bm{k}\cdot \bm{p}$  Hamiltonian to describe the hole pocket of the WSe$_2$ layer and the electron pocket of the MoTe$_2$ layer (Fig.~\ref{fig:fig1}e). Importantly, due to the nearby moir\'e bands, the electron pocket is described by a massive Dirac Hamiltonian which plays an important role in achieving an order parameter with $p_x+ip_y$ dependence which gives rise to a TIVC state.

\begin{figure}
		\centering
		\includegraphics[width=0.7\linewidth]{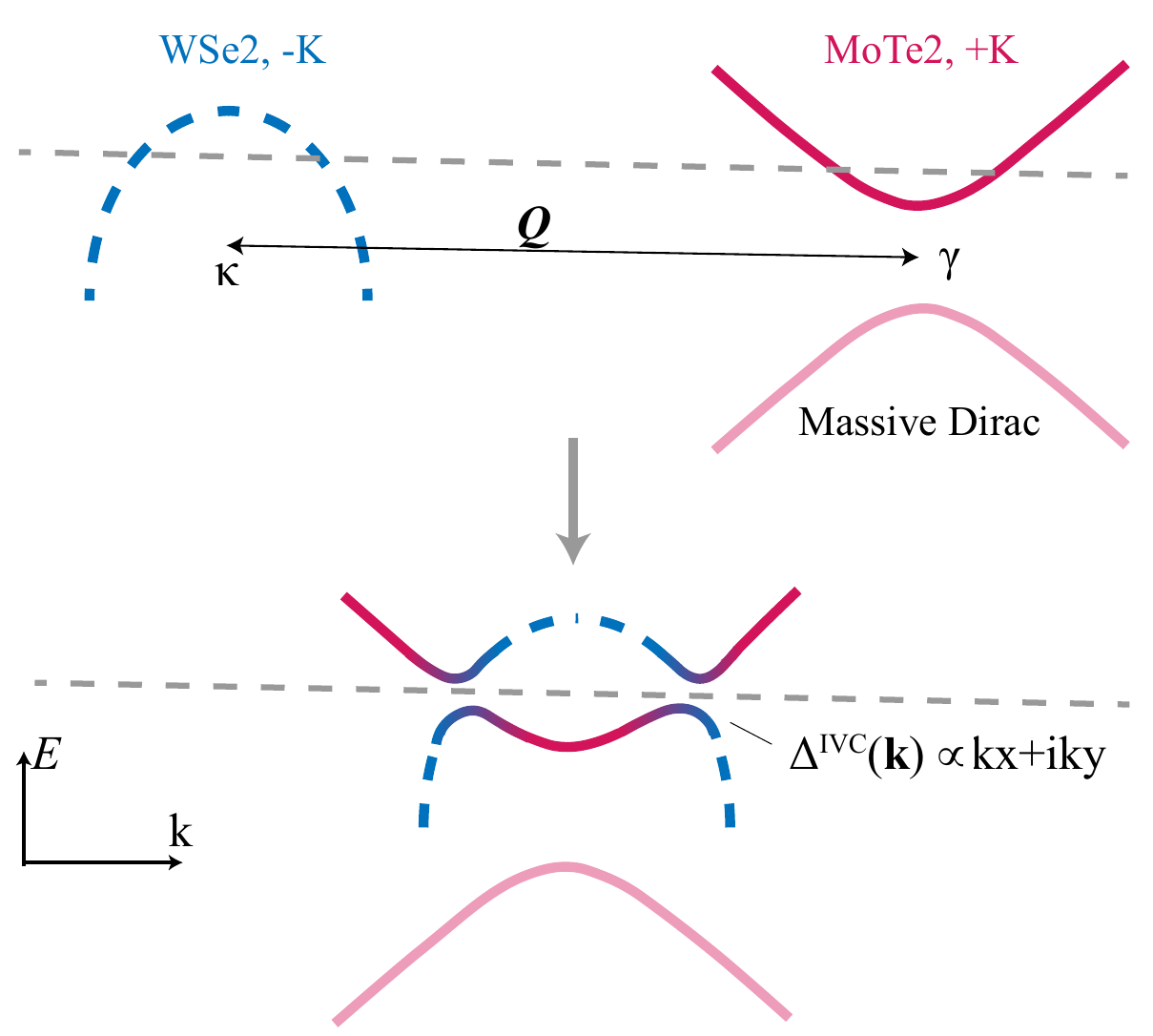}
		\caption{The top panel is a schematic plot of the low energy band near half-filling as shown in Fig.~\ref{fig:fig1}d. It constitutes of a hole band arising from the  moir\'e band top of the WSe$_2$ layer near $\kappa$ point of the $-K$ valley, and an electron band arising from the top moir\'e bands near the $\gamma$ point of the $K$ valley.  $\bm{Q}$ is the vector connecting $\gamma$ and $\kappa$ in the moir\'e Brillouin zone. The bottom channel shows that a topological insulating gap $ \Delta^{IVC}(\bm{k})  \propto k_x+ik_y$ can be opened in the TIVC state.}
		\label{fig:fig2}
	\end{figure}

	As shown in Figs.~\ref{fig:fig1}d and \ref{fig:fig1}e, in the presence of spin polarization, the low energy physics of the heterobilayer is governed by the hole pocket of the WSe$_2$ layer near the $\kappa$ point and the electron pocket of the MoTe$_2$ layer near the $\gamma$ point of the moir\'e Brillouin zone.   The $\kappa$ pocket representing the valence band top of WSe$_2$ layer at $-K$ is far away from other bands and it can simply be described by the band dispersion of $\xi_{-}(\bm{k})=-\bm{k}^2/2m_t-\mu+\delta E_D$ with $\mu$ as the chemical potential. Note that $\delta E_D$ has the same meaning as the $\delta E_D$ in Eq.~\ref{continuum_model}  but with a constant shift.  On the other hand, the top two moir\'e bands near the $\gamma$ pocket of $+K$ valley are close in energy and they are captured by a $C_{3v}$ symmetry invariant massive Dirac Hamiltonian  
	\begin{equation}
		H_{+,\gamma}(\bm{k})=\begin{pmatrix}
			\epsilon_{0}(k)+\Delta_M(k)&-v_F(k_x+ik_y)\\
			-v_F(k_x-ik_y)&\epsilon_{0}(k)-\Delta_M(k)
		\end{pmatrix},\label{massive_Dirac}
	\end{equation}
	where $\epsilon_{0}(k) = A_{0}(k_{x}^2+k_{y}^2)-\mu$, $\Delta_M(k)=\Delta_{M}+B_{0}(k_{x}^2+k_{y}^2)$. Here, $\Delta_M$ is the Dirac mass that is determined by the moir\'e potential. 
	To be specific, we  set $v_F=448 \; \rm{meV\cdot \AA}$, $A_{0}=3903 \; \rm{meV\cdot \AA^2}$, and $B_{0}=-758 \; \rm{meV\cdot \AA^2}$, being estimated from the moir\'e bands in Fig.~\ref{fig:fig1}.   The largest eigenenergies of $H_{\gamma}(\bm{k})$ is $\xi_{+}(\bm{k})= \epsilon_{0}(k)+\sqrt{v_F^2k^2+\Delta^2_M(k)}$, capturing the band dispersion of the spin-polarized bands near the $\gamma$  pocket. 
	The three-band low-energy effective  Hamiltonian  which describes the states near the Fermi energy can thus be written as 
	\begin{equation}
		H^{eff}_0(\bm{k})=\text{diag}[	H_{+,\gamma}(\bm{k}), \xi_{-}(\bm{k})]. \label{H_eff}
	\end{equation}
	Importantly, we will show that due to the massive Dirac dispersion near the  moir\'e band minimum of MoTe$_2$ layer ($\gamma$ pocket), the resulting state exhibits a $p+ip$ topological gap [see a schematic illustration in Fig.~\ref{fig:fig2}]. 
	
	\section{TIVC order parameter} 
	
	Given the electron and hole Fermi pockets in Fig.~\ref{fig:fig1}e, a natural way to obtain a fully gapped state is through hybridizing the states of two valleys via the inter-valley density-density interaction at half-filling as schematically shown in Fig.~\ref{fig:fig2}]. This is a classic scenario to form a finite momentum excitonic state \cite{HALPERIN1968115}. A distinct and important ingredient here is that the electron Fermi pocket arises from the conduction band of a massive Dirac. In this section, we demonstrate how the massive Dirac Hamiltonian of the MoTe$_2$ moir\'e bands and Coulomb interactions would naturally give rise to a finite momentum topological QAH state.   
	
	The interacting Hamiltonian can be written as 
	\begin{eqnarray}
		H_{int}=\frac{1}{2A}\sum_{\tau\tau'}\int  d\bm{r}d\bm{r'} V(\bm{r}-\bm{r'}):\rho^{\tau}(\bm{\bm{r}})\rho^{\tau'}(\bm{\bm{r'}}):, 
	\end{eqnarray}
	where $A$ is the sample area, $V(\bm{r}-\bm{r'})$ denotes Coulomb interaction,  the charge density operator $\rho^{\tau}(\bm{r})=\sum_{\sigma }c^{\dagger}_{\tau\sigma}(\bm{r})c_{\tau\sigma}(\bm{r})$ with $\sigma$ as spin indices. Note that the layer indices are locked with the spin indices near the Fermi energy due to the giant Ising spin-orbit coupling \cite{XiaoDi2012,Lu2015,Xi2016}.  Next, we perform a Hartree-Fock mean-field approximation to the interacting Hamiltonian $H_{int}$  and obtain the TIVC state at half-filling indicated by the black dashed line in Fig.~\ref{fig:fig1}d. The detailed formulation is presented in Appendix C and briefly sketched here. 
	
	We first project the interaction  onto the top two  moir\'e bands $\xi_{+,\bm{k}}$ and $\xi_{-,\bm{k}}$, 
	given that only these two moir\'e bands cross the Fermi energy.  Then we perform the Hartree-Fock mean-field approximation by taking a spin-polarized ground state ansatz as $\ket{\Phi}=\Pi_{\bm{k}}[u_{\bm{k}}c^{\dagger}_{+}(\bm{k})+v_{\bm{k}}c^{\dagger}_{-}(\bm{k}+\bm{Q})]\ket{0}$ with $|u_{\bm{k}}|^2+|v_{\bm{k}}|^2=1$, where $c^{\dagger}_{+}(\bm{k})$ and $c^{\dagger}_{-}(\bm{k}+\bm{Q})$ are the single-particle electron creation operators near the $\gamma$-pocket of MoTe$_2$ and the $\kappa$-pocket of WSe$_2$ layer with the same spin (Fig.~\ref{fig:fig1}d).  We expect the largest hybridization through the Coulomb interaction to appear when $\bm{k}$ and $\bm{k}+\bm{Q}$ are near the Fermi pockets of two valleys (see Fig.~\ref{fig:fig1}e), where $\bm{k}$ is now close to zero, $\bm{Q}$ is a vector connecting the $\kappa$  and the $\gamma$ point. The inter-valley coherent (IVC) order parameter is related to $\braket{\Phi|c^{\dagger}_{+}(\bm{k})c_{-}(\bm{k}+\bm{Q})|\Phi}$, which breaks the valley $U_{v}(1)$ symmetry and can be calculated self-consistently. The resulting self-consistent gap equation  for the IVC state is obtained as 
	
	\begin{equation}
		\Delta_{IVC}(\bm{k})=\frac{1}{2}\sum_{\bm{k'}}\tilde{V}^{+-}_{\bm{k',\bm{k}+\bm{Q},\bm{k}-\bm{k'}}}\chi(\bm{k'},\bm{Q})\Delta_{IVC}(\bm{k'}).\label{sq_1}
	\end{equation}
	Here,  $\chi(\bm{k'},\bm{Q})$ is the susceptibility function (see the Method section for its specific form). The effective interaction is given by $\tilde{V}^{+-}_{\bm{k',\bm{k}+\bm{Q},\bm{k}-\bm{k'}}}= \sum_{\bm{G},\bm{G'}}	V_{\bm{G}\bm{G'}}(\bm{k}-\bm{k'})\Lambda_{\bm{G}}^{+}(\bm{k},\bm{k'})\Lambda_{\bm{G'}}^{-}(\bm{k'}+\bm{Q},\bm{k}+\bm{Q})$, where  $	V_{\bm{G}\bm{G'}}(\bm{k}-\bm{k'})$ is the Fourier component of Coulomb interaction $V(\bm{r}-\bm{r'})$. The form factor is defined as $\Lambda_{\bm{G}}^{\tau}(\bm{k},\bm{k'})=\braket{\bm{k},\tau|e^{i(\bm{\bm{k}-\bm{k'}}+\bm{G})\cdot\bm{r}}|\bm{k'},\tau}$ with $\bm{G}$ as the moir\'e reciprocal lattice vectors, $\ket{\bm{k},\tau}$ as the Bloch wavefunction of top moir\'e band at $\tau$ valley. The form factor characterizes how the bare interaction is effectively dressed after projecting onto the band basis.
	
	
	Based on this self-consistent gap equation, we can classify the  $\Delta_{IVC}(\bm{k})$ using the point group symmetry of MoTe$_2$/WSe$_2$ heterobilayer. Under $C_{3z}$ operation, $\Delta_{IVC}(\bm{k})$  can be decomposed into $A$- and $E$-irreducible representations, which correspond to   the s-wave channel $\Delta_{IVC,A}=\Delta_0$ and the $p$-wave  channel $
	\Delta_{IVC,E}^{\pm}=\Delta_0 (k_x\pm ik_y)$, respectively. Note that if the $C_3$ symmetry is not broken in the ground state, these two channels would not mix.   Importantly,  when $\Delta_{IVC}(\bm{k})$ is dominated by $\Delta_{E,+}$ or  $\Delta_{E,-}$, the order parameter is a chiral  $p_x\pm ip_y$ wave and thus would be topologically nontrivial, being analogous to the $p_x\pm ip_y$ superconductor with a particle-hole transformation. Moreover in our case, the $p_x+ip_y$ and $p_x-ip_y$ IVC states are locked with the spontaneously spin polarization directions as required by time-reversal operation, and an external magnetic field can thus drive a transition between these two.
	
	It is also important to note that the momentum dependence of the IVC order parameter  	$\Delta_{IVC}(\bm{k})$ is now all encoded  in 
	the effective interaction $\tilde{V}^{+-}_{\bm{k',\bm{k}+\bm{Q},\bm{k}-\bm{k'}}}$ ($\bm{k'}$ is summed over). Specifically, it can be seen that the $\bm{k}$-dependence of the effective interaction is included in the interaction strength  $V_{\bm{G}\bm{G'}}(\bm{k}-\bm{k'})$ and the form factors $\Lambda_{\bm{G}}^{+}(\bm{k},\bm{k'})\Lambda_{\bm{G'}}^{-}(\bm{k'}+\bm{Q},\bm{k}+\bm{Q})$. On the other hand, the phase information is enclosed in the form factors as the interaction strength is a positive real quantity. The form factor near the moir\'e band top of WSe$_2$ labeled by $\kappa$ pocket of $-K$ valley in Fig.~\ref{fig:fig1}e is simply a plane wave so that $\Lambda_{\bm{G'}}^{-}(\bm{k'}+\bm{Q},\bm{k}+\bm{Q})=\delta_{\bm{G'},0}$. However, the form factor of the moir\'e band bottom of MoTe$_2$ at the $\gamma$ valley can be nontrivial. 
	
	Next, we point out that the massive Dirac behavior near $\gamma$ pocket can give rise to the nontrivial form factor needed for the $p+ip$ IVC order parameter, as schematically plotted in Fig.~\ref{fig:fig2}.
	The symmetry invariant two-band massive Dirac Hamiltonian near $\gamma$ point has been shown in Eq.~\ref{massive_Dirac}. The corresponding eigenstate  of top moir\'e band  $\xi_{+}(\bm{k})$ is  $\ket{\bm{k},\tau=+1}=(\cos\frac{\theta_{k}}{2},-\sin\frac{\theta_{k}}{2}e^{i \varphi_{\bm{k}}})^{T}$.   Here, $\theta_{k}=\arctan[\frac{v_F k}{\Delta_M(k)}]$ with $k=|\bm{k}|$,  $\varphi_{\bm{k}}=\text{Arg}[k_x+ik_y]$.
	This eigenstate can be expressed in the plane wave basis to evaluate the form factor.  Note that the form factor obtained from the Dirac model and the original continuum model shows a consistent behavior (see Fig.~\ref{FIG_S1} in Appendix D). The dominant contribution to the effective interaction comes from the  form factor $	\Lambda^{+}_{\bm{G}=0}(\bm{k},\bm{k'})$ assuming  that the interaction strength  $V(\bm{q})$ peaks at around a small $\bm{q}$, which can be obtained as
	\begin{equation}
		\Lambda^{+}_{\bm{G}=0}(\bm{k},\bm{k'})=\cos\frac{\theta_{k}}{2}\cos\frac{\theta_{k'}}{2}+\sin\frac{\theta_{k}}{2}\sin\frac{\theta_{k'}}{2}e^{i(\varphi_{\bm{k}}-\varphi_{\bm{k'}})}.	\label{form_factors}
	\end{equation}
	From the momentum dependence of the two terms in Eq.~\ref{form_factors}, we infer that the first term is an $s$-wave IVC (s-IVC) channel, while the second term is the $p_x+ip_y$ IVC. 	Importantly, as the form factor exhibits an angular dependence phase $e^{i\varphi_{\bm{k}}}$, the resulting IVC order parameter in general could generate a $p_x+ip_y$-wave channel.
	
	To show the $p_x+ip_y$  IVC states explicitly, we substitute the form factors back to the self-consistent gap equation Eq.~(\ref{sq_1}) and obtain
	\begin{align}
		\Delta_{IVC}(\bm{k})&=\cos\frac{\theta_{k}}{2}\sum_{\bm{k'}}V_0(\bm{k}-\bm{k'})F_1(\bm{k'})\nonumber\\&+\sin\frac{\theta_{k}}{2}e^{i\varphi_{\bm{k}}}\sum_{\bm{k'}}V_0(\bm{k}-\bm{k'})F_2(\bm{k'}),\label{sqq_2}
	\end{align}
	where $F_1(\bm{k'})=\frac{1}{2} \cos\frac{\theta_{k'}}{2}e^{-i\varphi_{\bm{k'}}}\chi(\bm{k'},\bm{Q})$, $F_2(\bm{k'})=\frac{1}{2} \sin\frac{\theta_{k'}}{2}\chi(\bm{k'},\bm{Q})$, $V_0(\bm{k}-\bm{k'})=\frac{2\pi e^2}{\epsilon\sqrt{|\bm{k}-\bm{k'}|^2+\lambda^2}}$ with $\epsilon$ as the dielectric constant, $\lambda^{-1}$ as an effective screening length. 
	
	We can decouple Eq.~\ref{sqq_2} into two self-consistent equations and solve $s$-IVC order parameter and $p_x+ip_y$-IVC order parameter independently by assuming the $C_3$ symmetry is not spontaneously broken. In the limit of $v_Fk_F\gg \Delta_M$ ($\theta_{k}$ is close to $\pi/2$) and weak interaction limit ($ V_0=\frac{2\pi e^2}{\epsilon\lambda}\ll \Lambda_c$, $\Lambda_c$ is a energy cutoff for the Dirac Hamiltonian), \YM{we can  obtain the $s$-IVC and TIVC order parameter analytically:
		\begin{eqnarray}
			\Delta^{s}_{IVC}(\bm{k})&&\approx\Delta_0\cos \frac{\theta_{k}}{2},\label{s_gap1}\\ 	\Delta^{p_x+ip_y}_{IVC}(\bm{k})&&\approx\Delta_0\sin \frac{\theta_{k}}{2} e^{i\varphi_{\bm{k}}} . \label{p_gap2}
		\end{eqnarray} 
		with $\Delta_0=2\Lambda_ce^{-\frac{1}{N_0\tilde{V_0}}}$.}
	Here,  $N_0$ is the average density of states near Fermi energy. The $\Delta_{IVC}(\bm{k})$ given by Eq.~(\ref{p_gap2}) is clearly  a $p_x+ip_y$  IVC  in analogous to the $p_x+ip_y$ pairing gap of a 2D spinless topological superconductor \cite{Alicea_2012}. Therefore, we have demonstrated how to obtain the TIVC state through a mean-field analysis. In the next section, we demonstrate the topological properties, including the calculations of the Chern number, through a full numerical calculation going beyond the weak coupling limit argument.
	
	\begin{figure*}
		\centering		\includegraphics[width=0.8\linewidth]{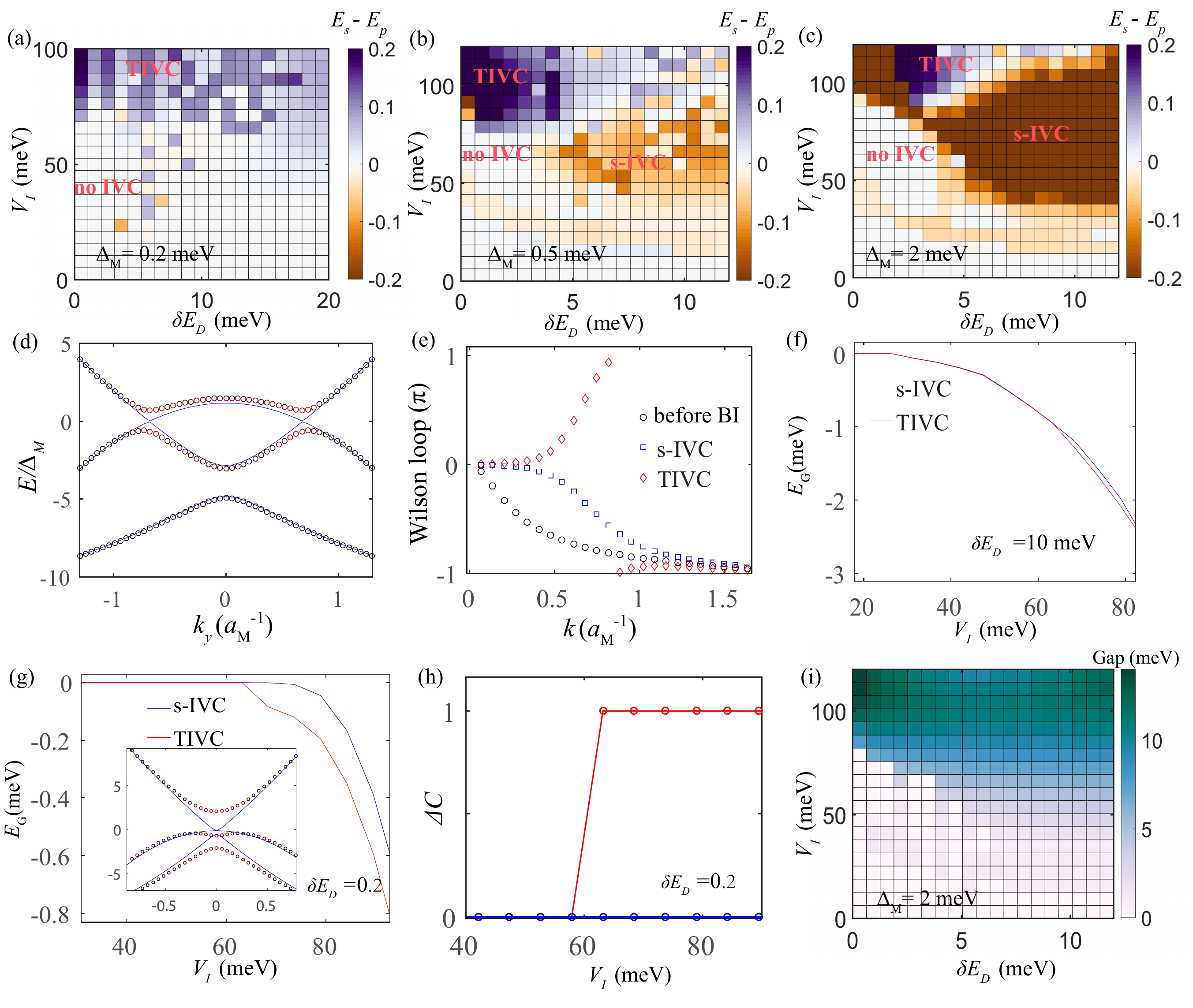}
		\caption{The TIVC state from the self-consistent Hartree-Fock calculation. (a)-(c) The ground-state energy difference of TIVC and $s$-IVC: $E_s-E_p$ as a function of $\delta E_D$ and interacting strength $V_I$ with a Dirac mass of $\Delta_M=0.2, 0.5, 2$ meV, respectively.  To be specific, we set the momentum cut-off as $1.3 a_M^{-1}$. The purple (brown) region indicates that TIVC (s-IVC) is more favorable.    (d) The quasiparticle energy spectrum (in units of $\Delta_M$) with an interaction-driven IVC gap (dotted lines), where the red color on the dispersion highlights the energy shift due to the IVC gap. For comparison, the noninteracting energy bands of the WSe$_2$ layer near the $\kappa$ point and the MoTe$_2$ layer near the $\gamma$ point    (blue lines) are also depicted. (e) The Wilsoon loop for the states of $s$-IVC, TIVC, before band inversion (BI).  (f), (g), respectively,
			shows the ground state energy (per electron) as a function of the interaction strength in a high-density limit ($\delta E_D=10$ meV)  and a low-density limit ($\delta E_D=0.2$ meV) with $\Delta_M=0.2$ meV. The inset shows the quasi-particle spectrum before (solid blue lines) and after (dotted lines) adding the Coulomb interaction ($V_I=70$ meV) in this case.  (h) The Chern number change $C$ of this system as a function of $V_I$ of the case shown in (g), where the red line corresponds to the TIVC states and the blue line corresponds to the $s$-IVC.  (i) The energy gap as a function of $V_I$ and $\delta E_D$ with $\Delta_M=2$ meV.   }
		\label{fig:fig3}
	\end{figure*}
	
	\section{ TIVC principle illustration from numerics}  
	To further illustrate the principle of TIVC,  we start with an effective mean-field  model Hamiltonian:
	$H_{MF}(\bm{k})=H^{eff}_0(\bm{k})+\Delta(\bm{k})$, where  $H^{eff}_0(\bm{k})$ is the three-band effective Hamiltonian given by Eq.~(\ref{H_eff}).  Here, $\Delta(\bm{k})$ is a three-by-three order parameter matrix with $\Delta_{ij}(\bm{k})=-\frac{1}{A}\sum_{\bm{k'}}\tilde{V_{ij}}(\bm{k}-\bm{k'})(\rho_{ij}(\bm{k})-\rho_0(\bm{k}))$, where   $\rho_{ij}(\bm{k})$ is the density matrix and $\rho_{0}(\bm{k})$ is the  noninteracting density matric with chemical potential $\mu$ at half-filling. For simplicity, we take $\tilde{V}_{ij}(\bm{k}-\bm{k'})=V_I/\sqrt{(\bm{k}-\bm{k'})^2+\lambda^2}$ where $V_I$ is a phenomenological parameter  characterizes the Coulomb interaction strength. This minimal three-band mean-field model captures the essential ingredients, including the massive Dirac spectrum and the Coulomb interactions of the heterobilayer. Note that a self-consistent with the TIVC state using full moir\'e bands is beyond the scope of this work, which would involve heavy numerics.

	\YM{
		We define the ground state energy as $E_G=E_t(V_I)-E_t(V_I=0)$, where the zero temperature total  energy  can be calculated at  given interaction $V_I$ with
		\begin{equation}
			E_t(V_I)=\frac{1}{N}\sum_{n,\bm{k}} E_{n,\bm{k}}.
		\end{equation}
		Here, $N$ is the total number $\bm{k}$ points, $n=\{2,3\}$ labels the two bands below the Fermi energy, $E_{n,\bm{k}}$ is the $n$-th band given by Hartree-Fock calculation.  The phase diagram can be obtained by comparing the ground state energy $E_G$ of different states in various parameter regions. Note that there would be no IVC  if $E_G>0$. In our model, the crucial parameters affecting the IVC order are the interaction strength $V_I$, gate potential $\delta E_D$, and the massive gap $\Delta_M$.  The increase of $\delta E_D$ enables the WSe$_2$ bands to overlap the top MoTe$_2$ band in energy. At $\delta E_D=0$, the valence band top of the WSe$_2$ band is at the middle of the massive gap of MoTe$_2$ bands.	One gauge invariant and efficient way to determine the topological properties of bands after Hartree-Fock calculation is to use the Wilson loop method \cite{Xue2018,Dai2011,Hongming2015}.  To distinguish the $s$-IVC state and TIVC state, we can calculate the Wilson loop along different square loops around the momentum space origin, which is equal to the integration of the Berry curvature over these momentum loops. }
	
	
	\YM{ By evaluating the energy difference between $s$-IVC and TIVC state ($E_s-E_p$), the phase diagrams in the $\delta E_D$-$V_I$ plane with a Dirac mass of $\Delta_M=0.2, 0.5, 2$ meV are shown in Fig.~\ref{fig:fig3}a-c, respectively.   These phase diagrams typically contain distinctive regions (note that some noise appears because the self-consistency is accidentally saturated in the metastable states). In the region without IVC, we find that Coulomb interaction is not strong enough to establish a finite order parameter. Note that we determine the existence of IVC states by checking whether $E_G$ is negative or not.  When Coulomb interaction is larger (tens of meV), the IVC order can start establishing in the higher density region. Typically,  a correlated insulating gap is opened near the band crossing points by the interactions (Fig.~\ref{fig:fig3}d). } 
	
	
	\YM{In the region with IVC, we clearly find two possible saturated gapped states after Hartree-Fock self-consistency: $s$-IVC and TIVC states. To show the distinct topology between $s$-IVC and TIVC states, the Wilson loops of the highest bands after Hartree-Fock calculation as a function of momentum length $\bm{k}$ (half of the side length of square loops) are shown in Fig.~\ref{fig:fig3}(e). When the WSe$_2$ band is far below the top band of MoTe$_2$ (before band inversion (BI)), the Wilson loop simply represents the massive Dirac band behavior with a winding of $-\pi$ (the black dot line, labeled as before band inversion (BI)).  After the WSe$_2$  band is pushed up by the gate, the Coulomb interaction is able to drive the IVC gap near the Fermi energy by interacting with the MoTe$_2$ bands. In this case, by using different initial input IVC order parameters in the Hatree-Fock calculation,  we find the Wilson loop of the resulting top Hartree-Fock band can either preserve a winding of $-\pi$ (the blue line) or change as a winding of $\pi$ (red dot line). The latter would imply that the Chern number changes one after opening the IVC gap representing the TIVC states, while the former IVC gap does not change the topology of this system representing the $s$-IVC states.}

	\YM{After showing how we distinguish the $s$-IVC and TIVC in the Hartree-Fock calculation, we discuss the energy difference between the TIVC and $s$-IVC  in more detail. We first focus on the small Dirac mass limit.  As an illustration,  line cuts of ground state energy $E_G$  in Fig.~\ref{fig:fig3}(a) as a function of the interaction strength at a higher density region $\delta E_D=10$ meV and a low-density region $\delta E_D=0.2$ meV are shown in Fig.~\ref{fig:fig3}f and Fig.~\ref{fig:fig3}g, respectively. In the high-density limit ($v_Fk_F\gg\Delta_M$), we find the energy of  $s$-IVC and TIVC are degenerate at smaller interaction region ($V_I\lesssim 60$ meV).  The ground state's energy should mostly be determined by the gap size of the order parameters. From this point of view, this degeneracy is expected by our analytical analysis (Eqs.~(\ref{s_gap1}) and (\ref{p_gap2})), where the $s$-IVC and $p_x+ip_y$ IVC possess the same gap size in the limit ($v_Fk_F\gg\Delta_M$). Furthermore, we observe a clear degeneracy lifting at a larger interaction strength region ($V_I> 60$ meV).  Importantly, such lifting is even more clear at low-density regions (Fig.~\ref{fig:fig3}g). Interestingly, we note that the essential physical ingredient that causes the TIVC to be more favorable than the $s$-IVC at large interaction limit arises from the valence band of the massive Dirac of MoTe$_2$. To illustrate this point,   the energy spectrum after the mean-field calculation with $V_I=80$ meV and $\delta E_D=0.2$ meV are shown in the inset of Fig.~\ref{fig:fig3}g. Our calculation suggests that in the small $\Delta_M$ limit,   the valence band of  MoTe$_2$ is more critical in helping to stabilize the TIVC parameters.}

	\YM{Now we discuss how our results are affected in a larger $\Delta_M$ limit (Fig.~\ref{fig:fig3}b-c).  We find the TIVC phase would be gradually suppressed by the increase of $\Delta_M$. One important reason is that the $s$-IVC gap would be larger than the TIVC gap away from the $v_F k_F\gg \Delta_M$ limit.  The other related reason is the increase of the massive gap would effectively weaken the effects of nontrivial form factors of massive Dirac. When $\Delta_M \sim v_F k_F$, the IVC states can be regarded as an excitonic gap between two quadratic bands which is expected to be $s$-wave.}

Therefore, our analysis unambiguously shows the existence of the TIVC states in theory. Our theory also suggests that such TIVC states would be more energy-favorable in a smaller $\Delta_M$ limit. On the other hand, the Dirac mass $\Delta_M$ at MoTe$_2$ layer is proportional to the moir\'e potential. It is also worth emphasizing that the analysis in this part mainly is to illustrates the principle of the TIVC state. The detailed values of parameters are not our concern here, which are hard to determine from purely theoretical analysis.


\YM{Before ending this section, we highlight how the quasi-particle gap changes in the $\delta E_D$-$V_I$ plane in Fig.~\ref{fig:fig3}i. Interestingly, we do not find gap-closing during the phase transition between  $s$-IVC and TIVC states, although the bulk topology between these two gaped states is different. It is in sharp contrast with non-interacting topological insulators, where the topological phase transition is driven by the gap-closing and re-opening. This is also consistent with the absence of bulk gap-closing in the experiment during entering the QAH state \cite{Fai_nature2021}. }

\section{Discussion}

To summarize, the new layer-resolved MCD experiments suggested that, near half-filling, the top moir\'e bands of the MoTe$_2$ layer and the WSe$_2$ layer would have the same spin polarization. Moreover, the QAH state with quantized Hall resistance would appear when the top moir\'e band of WSe$_2$ is hole-doped. The new experiments suggest that the band structure at half-filling has the form shown in Fig.~\ref{fig:fig1}b. However, how a QAH state can emerge out of the band structure shown in Fig.~\ref{fig:fig1}b was not known. In this work, based on the assumption that a Hund's splitting can push the WSe$_2$ band up to the Fermi energy, we have established a theory that can explain many of the key experimental observations. Specifically, we have demonstrated how the Coulomb interactions can couple the states from the hole pockets and the electron pockets of the WSe$_2$ layers and the MoTe$_2$ layers respectively, which are separated by a constant momentum vector $\bm{Q}=(\frac{4\pi}{3a_M},0)$. Importantly, we have shown the massive Dirac dispersion nature of the MoTe$_2$ moir\'e bands provides a phase winding in the order parameter to give rise to a QAH state. Further self-consistent Hartree-Fock calculations confirm that the top moir\'e bands carry finite Chern numbers for a certain range of Coulomb interaction strengths and gating as observed in the experiment.

\begin{figure*}
\centering
\includegraphics[width=0.8\linewidth]{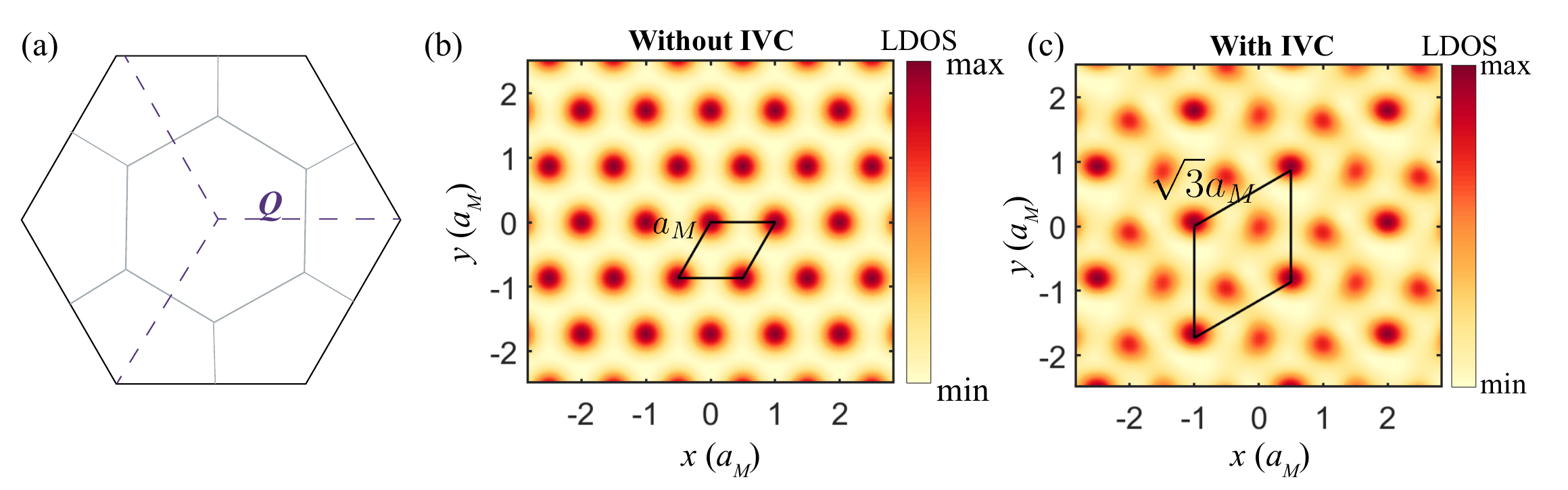}
\caption{The charge density wave behavior induced by the IVC state with the finite momentum $\bm{Q}=(\frac{4\pi}{3a_M},0)$. (a) The folded Brillouin zones (gray lines) due to the finite $\bm{Q}$, which are one-third of the original moir\'e Brillouin zone (black line). (b) and (c), respectively, show the local density of state (LDOS)  $\rho(E,\bm{r})$ ($E$ is set near half-filling) for the case without ($\Delta_{IVC}=0$) and with IVC state ($\Delta_{IVC}=1$ meV), in which the $a_M\times a_M$ moir\'e unit cell and the enlarged $\sqrt{3}a_M\times \sqrt{3}a_M$  unit cell are highlighted. }
\label{fig:fig4}
\end{figure*}

We would like to emphasize that the hybridization of the states with a constant momentum vector $\bm{Q}$ has an important and interesting experimental consequence. Due to the hybridization, the moir\'e bands are further folded such that the $\bm{Q}$-vector becomes a new reciprocal vector of the heterobilayer. As a result, the unit cells are enlarged and a new unit cell is induced by the order parameter $\braket{c^{\dagger}_{+,\bm{k}}c_{-,\bm{k}+\bm{Q}}}$. The local density of states in the absence and presence of the TIVC order parameter is shown in Fig.~\ref{fig:fig4}. It is clearly shown that the new unit cell is three times as large as the original moir\'e unit cell. The predicted change in the local density of states can be verified experimentally through scanning tunneling microscopy (STM) \cite{Zalatel2022}.  \YM{Recently, the IVC states were experimentally studied in twisted multilayer graphene with STM measurements \cite{Nuckolls2023,Kim2023}.  }

To conclude, our theory suggests that the QAH state observed in the recent experiment \cite{Fai_nature2021, Faitalk} is an interesting TIVC state with finite momentum pairing. This state is a new state of matter that is different from the previously proposed QAH state in magnetic doped topological insulators and valley-polarized QAH states in twisted bilayer graphene. This work provides insights for further exploring the novel properties of the observed unconventional QAH state in MoTe$_2$/WSe$_2$ heterbilayers. \YM{  Our work would also  motivate the experimental exploration of the IVC or excitonic states with nontrivial topology in moir\'e materials. Finally,   how the TIVC state would interplay with Hubbard physics could be an interesting direction to explore in the future \cite{Chen2022}. } 

\bigskip

\section*{Acknowledgments} 

The authors thank Kin Fai Mak for showing us the MCD experimental data prior to its publication and thank him for insightful discussions. We also thank Adrian H.C. Po for inspiring discussions. K.T.L. acknowledges the support of the Ministry of Science and Technology, China, and HKRGC through Grants No. 2020YFA0309600, No. RFS2021-6S03, No. C6025-19G, No. AoE/P-701/20, No. 16310520, No. 16310219, No. 16307622 and No. 16309718. Y.M.X. acknowledges the support of HKRGC through Grant No. PDFS2223-6S01.




\begin{appendix}

\section{Details for the continuum model Hamiltonian}
The continuum model	
$\mathcal{H}_{\tau}(\bm{r})=
\begin{pmatrix}
\mathcal{H}^{b}_{\tau}(\bm{r})+\tau M_z
& W_{\tau}(\bm{r})\\
W_{\tau}^{\dagger}(\bm{r})
& \mathcal{H}^{t}_{\tau}(\bm{r}) -\tau M_z- \delta E_{D}
\end{pmatrix}$
is diagonalized under the plane wave basis functions $\{\ket{\bm{k}+\bm{G}}\}$, where $\bm{k}$ is defined in the moir\'e Brillouin zone, $\bm{G}=m\bm{G_2}+n\bm{G_3}$ are the reciprocal vectors for the moir\'e supercell, with $m$, $n$ being integer numbers and $\bm{G}_{j}=\frac{4\pi}{\sqrt{3}a_{M}}[-\sin \frac{(j-1)\pi}{3},\cos \frac{(j-1)\pi}{3}]$ defined in the main text. Within this basis, the representation of the Hamiltonian reads
\begin{align}
\mathcal{H}^{l'l}_{\bm{k}}(\bm{G'}, \bm{G})=\bra{\bm{k}+\bm{G'}}\mathcal{H}^{l'l}_{\tau}(\bm{r})\ket{\bm{k}+\bm{G}},
\end{align}
where $l', l$ are the layer indices. Here, we list the explicit forms of momentum, moir\'e potential, and interlayer hopping as follows
\begin{align}
\bra{\bm{k}+\bm{G'}}\hat{\bm{p}}\ket{\bm{k}+\bm{G}}=(\bm{k}+\bm{G})\delta_{\bm{G'},\bm{G}},\\
\bra{\bm{k}+\bm{G'}}V_{l}(\bm{r})\ket{\bm{k}+\bm{G}}=V_{l}(\bm{G'}-\bm{G}),\\
\bra{\bm{k}+\bm{G'}}W_{\tau}(\bm{r})\ket{\bm{k}+\bm{G}}=W_{\tau}(\bm{G'}-\bm{G}),
\end{align}
where $V_{l}(\bm{G'}-\bm{G})=V_{l}e^{i\phi_{l}}\sum\limits_{j=1,3,5}\delta_{\bm{G'}-\bm{G},\bm{G}_{j}}+V_{l}e^{-i\phi_{l}}\sum\limits_{j=2,4,6}\delta_{\bm{G'}-\bm{G},\bm{G}_{j}}$, and $W_{\tau}(\bm{G'}-\bm{G})= W(\delta_{\bm{G'},\bm{G}}+\omega^{\tau}\delta_{\bm{G'}-\bm{G},-\bm{G}_{2}}+\omega^{2\tau}\delta_{\bm{G'}-\bm{G},-\bm{G}_{3}})$ are the Fourier transforms of the moir\'e potential $V_{l}(\bm{r})$ and the interlayer hopping $W_{\tau}(\bm{r})$ respectively.

Using the above relations, we obtain the Hamiltonian in the momentum space with a momentum cut-off $-N\leq m,n \leq N$, where we take $N=2$. With the parameters listed in  Table.~\ref{table:S01}, we get the band structure of moir\'e MoTe$_2$/WSe$_2$ heterobilayers as shown in the main text. 

\begin{center}
\begin{table}[ht]
	\caption{Parameters for the continuum model.} 
	\centering 
	\begin{tabular}{c c c c c c c c} 
		\hline\hline 
		$m_{b} / m_{0}$ &  $m_{t} / m_{0}$ &  $\lambda_{b} \kappa ^{3}(\mathrm{meV})$ & 
		$V_{b}(\mathrm{meV})$ &  $\phi_{b}$ & \hspace{0.2 in}
		$W(\mathrm{meV})$ \\  
		0.65 &  0.35 &  5 &  4.1 &  14$\degree$ &  1.3 \\
		\hline\hline 
	\end{tabular}
	\label{table:S01} 
\end{table}
\end{center}

\section{The Hartree-Fock calculation for the moir\'e Hamiltonian in the plane wave basis} \YM{In the Fig.~\ref{fig:fig5}, we have shown how the valley polarization effects could generate the desired large Hund's splitting. Here, we present the details of how the valley-polarized bands in Fig.~\ref{fig:fig5} are obtained. The bands are obtained by diagonalizing the Hartree-Fock mean-field Hamiltonian, which  in the plane wave basis is given by
\begin{equation}
	H_{MF}=H_0+\Sigma^{H}+\Sigma^{F}.
\end{equation} 
Here, $H_0$ is the non-interacting moir\'e Hamiltonian,  the Hartree order parameter
\begin{equation}
	\Sigma^{H}_{\alpha\bm{G},\beta\bm{G'}}=\frac{1}{A}\sum_{\bm{k'}, \bm{G''}, \alpha'} V_{\bm{G'}-\bm{G}}  \delta \rho_{\alpha'\bm{G''}+\bm{G'}-\bm{G},\alpha' \bm{G'}}(\bm{k'}),
\end{equation}
where $\alpha, \beta$ is the internal indices including the layer, valley, and spin, $\bm{G}$ is the moir\'e reciprocal lattice vectors.  The bare Coulomb interaction $V_q=\frac{e^2}{4\pi\epsilon \sqrt{q^2+\lambda^2}}$, where $\epsilon$ is the dielectric constant, $\lambda^{-1}$ is the screening length.
The Fock order parameter
\begin{equation}
	\Sigma^{F}_{\alpha \bm{G},\beta\bm{G'}}(\bm{k})=-\frac{1}{A} \sum_{\bm{G''},\bm{k'}} V_{\bm{k'}-\bm{k}+\bm{G''}} \delta \rho_{\alpha \bm{G'}+\bm{G''};\beta\bm{G}+\bm{G''}}(\bm{k'}) 
\end{equation}
where $	\delta \rho_{\alpha \bm{G},\beta\bm{G'}}(\bm{k})=\rho_{\alpha \bm{G},\beta\bm{G'}}(\bm{k})-\rho_0$, and the density matrix
\begin{eqnarray}
	\rho_{\alpha \bm{G},\beta\bm{G'}}(\bm{k})=\braket{c^{\dagger}_{\alpha \bm{G}}(\bm{k}) c_{\beta\bm{G'}}(\bm{k})}\nonumber&&\\=\sum_{n} U^{n}_{\beta \bm{G'}}(\bm{k})U^{*n}_{\alpha \bm{G}}(\bm{k}) f(E_n(\bm{k})) 
\end{eqnarray}
with $\ket{\psi_{n\alpha}(\bm{k})}=\sum_{n,\bm{G}} U^*_{\alpha,\bm{G}}(\bm{k})\ket{c_{\alpha,\bm{G}}(\bm{k})}$, $\rho_0$ is the density matrix in non-interacting limit. The density matrix $\delta\rho$ is defined relative to the bare bands since the Hartree and Fock energy should vanish in the non-interacting limit. }

\section{The mean-field analysis for TIVC states with finite Q}

To see the effects of interactions clearly, we project the density-density interaction onto the two  moir\'e bands $E_{\tau}(\bm{k})$ near half-filling as shown in Fig.~\ref{fig:fig1}d,  arising from the $K$ valley of MoTe$_2$ layer and the $-K$ valley of WSe$_2$ layer with the same spin.  The effective Hamiltonian becomes
\begin{equation}
H_{eff}=\sum_{\bm{k},\tau}\xi_{\tau}(\bm{k})+\frac{1}{2A}\sum_{\bm{q},\bm{G}\bm{G'}}V_{\bm{G}\bm{G'}}(\bm{q}):\rho_{\bm{q+\bm{G}}}\rho_{-\bm{q}-\bm{G'}}:.
\end{equation} 
Here, $A$ is the sample area, $\xi_{\tau}(\bm{k})=E_{\tau}(\bm{k})-\mu$ denotes the two  moir\'e bands near Fermi energy with $\mu$ representing the chemical potential, the Fourier component of interaction $V_{\bm{G}\bm{G'}}(\bm{q})=\frac{2\pi e^2}{\epsilon\sqrt{|\bm{q}+\bm{G}||\bm{q}+\bm{G'}|+\lambda^2}}$.  The density operator $\rho_{\bm{G}}^{\tau}(\bm{q})=\sum_{\bm{k}}\Lambda_{\bm{G}}^{\tau}(\bm{k}+\bm{q},\bm{k})c^{\dagger}_{\tau}(\bm{k}+\bm{q})c_{\tau}(\bm{k})$, where we define the form factor $\Lambda_{\bm{G}}^{\tau}(\bm{k}+\bm{q},\bm{k})=\braket{\bm{k}+\bm{q},\tau|e^{i(\bm{q}+\bm{G})\cdot\bm{r}}|\bm{k},\tau}$.  
Then we expand  $H_{eff}$ in a Hartree-Fock mean field manner:
\begin{align}
H_t&=\sum_{\bm{k},\tau}\xi_{\tau}(\bm{k})c^{\dagger}_{\tau}(\bm{k})c_{\tau}(\bm{k})+\frac{1}{S}\sum_{\bm{k},\bm{q}}\sum_{\tau\tau'}(\Delta^{H}_{\tau\tau}(\bm{k},\bm{q})\delta_{\tau,\tau'}\nonumber\\
&+\Delta^{F}_{\tau\tau'}(\bm{k},\bm{q}))c_{\tau}^{\dagger}(\bm{k}+\bm{q})c_{\tau'}(\bm{k}).
\end{align}
Here, the Hartree and Fock order parameter is defined as
\begin{align}
&\Delta^{H}_{\tau\tau}(\bm{k},\bm{q})=\sum_{\bm{k'},\tau'}\tilde{V}_{\bm{k}\bm{k'}\bm{q}}^{\tau\tau'}\braket{c^{\dagger}_{\tau'}(\bm{k'}-\bm{q})c_{\tau'}(\bm{k'})},\\
&\Delta^{F}_{\tau\tau'}(\bm{k},\bm{q})=-\sum_{\bm{k'}}\tilde{V}^{\tau\tau'}_{\bm{k'},\bm{k},\bm{q}-\bm{k'}+\bm{k}}\braket{c^{\dagger}_{\tau'}(\bm{k'}-\bm{q})c_{\tau}(\bm{k'})}.
\end{align}

As we mentioned in the main text, we perform a mean-field approximation by taking the ground state ansatz at half-filling as $\ket{\Phi}=\Pi_{\bm{k}}[u_{\bm{k}}c^{\dagger}_{+}(\bm{k})+v_{\bm{k}}c^{\dagger}_{-}(\bm{k}+\bm{Q})]\ket{0}$ with $|u_{\bm{k}}|^2+|v_{\bm{k}}|^2=1$.   In this case, the only non-vanishing terms are
\begin{align}
\braket{\Phi|c^{\dagger}_{+}(\bm{k})c_{+}(\bm{k})|\Phi}=|u_{\bm{k}}|^2,\\\ \braket{\Phi|c^{\dagger}_{-}(\bm{k}+\bm{Q})c_{-}(\bm{k}+\bm{Q})|\Phi}=|v_{\bm{k}}|^2,\\
\braket{\Phi|c^{\dagger}_{-}(\bm{k}+\bm{Q})c_{+}(\bm{k})|\Phi}=u_{\bm{k}}v_{\bm{k}}^*,\\	\braket{\Phi|c^{\dagger}_{+}(\bm{k})c_{-}(\bm{k}+\bm{Q})|\Phi}=u^*_{\bm{k}}v_{\bm{k}}.
\end{align} 

After calculating the Hartree and Fock order parameter with these averages, the resulting effective mean field Hamiltonian is found to be
\begin{equation}
H^{eff}_{MF}=\frac{1}{A}\sum_{\bm{k}}\Psi^{\dagger}_{\bm{k},\bm{Q}}\begin{pmatrix}
	\tilde{\xi}_{+}(\bm{k})&\Delta_{IVC}(\bm{k})\\\Delta^*_{IVC}(\bm{k})
	&\tilde{\xi}_{-}(\bm{k})
\end{pmatrix}\Psi_{\bm{k},\bm{Q}},
\end{equation}
Here, the two-component creation operator is $\Psi^{\dagger}_{\bm{k},\bm{Q}}=(	c^{\dagger}_{+}(\bm{k}), 	c^{\dagger}_{-}(\bm{k}+\bm{Q}))^{T}$,  $\tilde{\xi}_{\tau}(\bm{k})$ is the interaction dressed moir\'e bands by including the Hartree and Fock energies into $\xi_{\tau}(\bm{k})$, the order parameter of the inter-valley coherent state is defined as $\Delta_{IVC}(\bm{k})$, which mixes the states of the maximum of the top moir\'e band of WSe$_2$ with momentum $\bm{k}$  and the minimum of the top moir\'e band of MoTe$_2$ with momentum $\bm{k}+\bm{Q}$. The quasi-particle annihilation operator for the filling states can thus be written as $
\gamma_{-}(\bm{k})=\sin\frac{\beta_{\bm{k}}}{2}e^{i\frac{\alpha_{\bm{k}}}{2}}c_{+}(\bm{k})+\cos\frac{\beta_{\bm{k}}}{2}e^{-i\frac{\alpha_{\bm{k}}}{2}}c_{-}(\bm{k}+\bm{Q})$ .  At half-filling, we can rewrite the gapped ground state as
\begin{equation}
\ket{\Psi}=\Pi_{\bm{k}}\gamma_{-}^{\dagger}\ket{0}.
\end{equation}  
The corresponding self-consistent equation is
\begin{equation}
\Delta_{IVC}(\bm{k})=\frac{1}{2}\sum_{\bm{k'}}\tilde{V}^{+-}_{\bm{k',\bm{k}+\bm{Q},\bm{k}-\bm{k'}}}\chi(\bm{k'},\bm{Q})\Delta_{IVC}(\bm{k'}). \label{Self_equation}
\end{equation}
Here, the susceptibility function $\chi(\bm{k'},\bm{Q})=(\xi'(\bm{k'})^2+|\Delta_{IVC}(\bm{k'})|^2)^{-1/2}$ with $\xi'(\bm{k'})=[\tilde{\xi}_{+}(\bm{k'})-\tilde{\xi}_{-}(\bm{k'}+\bm{Q})]/2$, and the effective interaction is given by
\begin{eqnarray}
&&\tilde{V}^{+-}_{\bm{k',\bm{k}+\bm{Q},\bm{k}-\bm{k'}}}= \sum_{\bm{G},\bm{G'}}	V_{\bm{G}\bm{G'}}(\bm{k}-\bm{k'})\times\nonumber\\&&\Lambda_{\bm{G}}^{+}(\bm{k},\bm{k'})\Lambda_{\bm{G'}}^{-}(\bm{k'}+\bm{Q},\bm{k}+\bm{Q}). \label{eff_interaction}
\end{eqnarray}

\section{Massive Dirac Hamiltonian near the top moir\'e band minimum}
we show that the effective Hamiltonian near the $\gamma$ pocket can also be derived by group theory, which can be easily generalized to include the higher order terms, and we further determine the coefficients by fitting the band structure obtained by the continuum model.

The folded bands at $\gamma$ point are given by $\epsilon_{i}= \epsilon_{b}+2V_{b}\cos (\phi_{b}+\frac{2\pi}{3}(i-1))$ \cite{Law_PRL2022}.  
In the basis of $(\ket{\epsilon_1},\ket{\CP{\epsilon_2}})$ (the closest bands near Fermi energy), the $C_3$ operation reads $D(C_3)=\begin{pmatrix}
1&0\\
0&\omega
\end{pmatrix}$, and the Pauli matrices transform as $\rho_{\pm} \rightarrow \CP{\omega^{\mp}} \rho_{\pm}$ under the $C_3$ operation with $\rho_{\pm}=\rho_{x}\pm \rho_{y}$, therefore belong to the $E$ representation. Similarly, the Pauli matrix $\rho_{z}$ belongs to the $A_{1}$ representation. We further list the irreducible representations of the momentum polynomials in Table.~\ref{table:S02}, and find the symmetry allowed effective Hamiltonian accordingly:
\begin{eqnarray}
H_{+,\gamma}(\bm{k})=\epsilon_{0}(k)-v_F(k_x\rho_x\CP{-}k_y\rho_y)\nonumber\\+\Delta_M(k)\rho_z+C_{0}[(k_{x}^2-k_{y}^2)\rho_x\CP{+}2k_xk_y\rho_y],
\end{eqnarray}
where $\epsilon_{0}(k) = A_{0}(k_{x}^2+k_{y}^2)-\mu$ and $\Delta_M(k)=\Delta_{M}+B_{0}(k_{x}^2+k_{y}^2)$. By fitting the bands obtained by the continuum model, we find the parameters $\Delta_{M}=\CP{5.07} \; \rm{meV}$, $v_F=448 \; \rm{meV\cdot \AA}$, $A_{0}=3903 \; \rm{meV\cdot \AA^2}$, $B_{0}=-758 \; \rm{meV\cdot \AA^2}$ and $C_{0}=6355 \; \rm{meV\cdot \AA^2}$. The $C_{0}$-term describes the warping effect, which is only used in our numerical calculations, but ignored in the analytical derivations for simplicity.

\begin{center}
\begin{table}[ht]
	\caption{Irreducible representations (IRs) of Pauli matrices and
		momentum under $3m'$ point group.} 
	\centering 
	\begin{tabular}{c c c c c c} 
		\hline\hline 
		IRs & Pauli matrices &  linear functions & 
		quadratic functions \\  
		\hline		
		A$_{1}$ &  $\rho_{z}$ & - &  $k_{x}^2+k_{y}^2$ \\			
		A$_{2}$ &  - &  - & 
		- \\ 
		E & $\{\rho_{x}, \; \rho_{y}\}$ &  $\{k_{x}, \; k_{y}\}$ & $\{k_{x}^2-k_{y}^2, \; 
		k_{x}k_{y}\}$ \\
		\hline\hline 
	\end{tabular}
	\label{table:S02} 
\end{table}
\end{center} 
To further validate the effectiveness of the massive Dirac model, we carry out a comparison of the form factors derived from the continuum and Dirac models, as shown in Fig.~\ref{FIG_S1}. For comparison, we examine our analytical solution of the form factor as will be derived in the Appendix F: $\Lambda^{+}_{\bm{G}=0}(\bm{k},\bm{k'})=\cos\frac{\theta_{k}}{2}\cos\frac{\theta_{k'}}{2}+\sin\frac{\theta_{k}}{2}\sin\frac{\theta_{k'}}{2}e^{i(\varphi_{\bm{k}}-\varphi_{\bm{k'}})}$. In order to plot the form factor, we fix $\bm{k'}=(|\bm{k}|,0)$ along the $x$ direction, and plot a 2D map with respect to $\bm{k}=(k_{x}, k_{y}) $. The color bar represents the modulus of the form factor $\Lambda^{+}_{\bm{G}=0}$, and the arrow indicates the phase. Specifically, the $x$ and $y$ components of the arrow represent the real and imaginary part of $\Lambda^{+}_{\bm{G}=0}/|\Lambda^{+}_{\bm{G}=0}|$ respectively show a good agreement with our analytical solution.

\begin{figure}[h]
\centering
\includegraphics[width=1\linewidth]{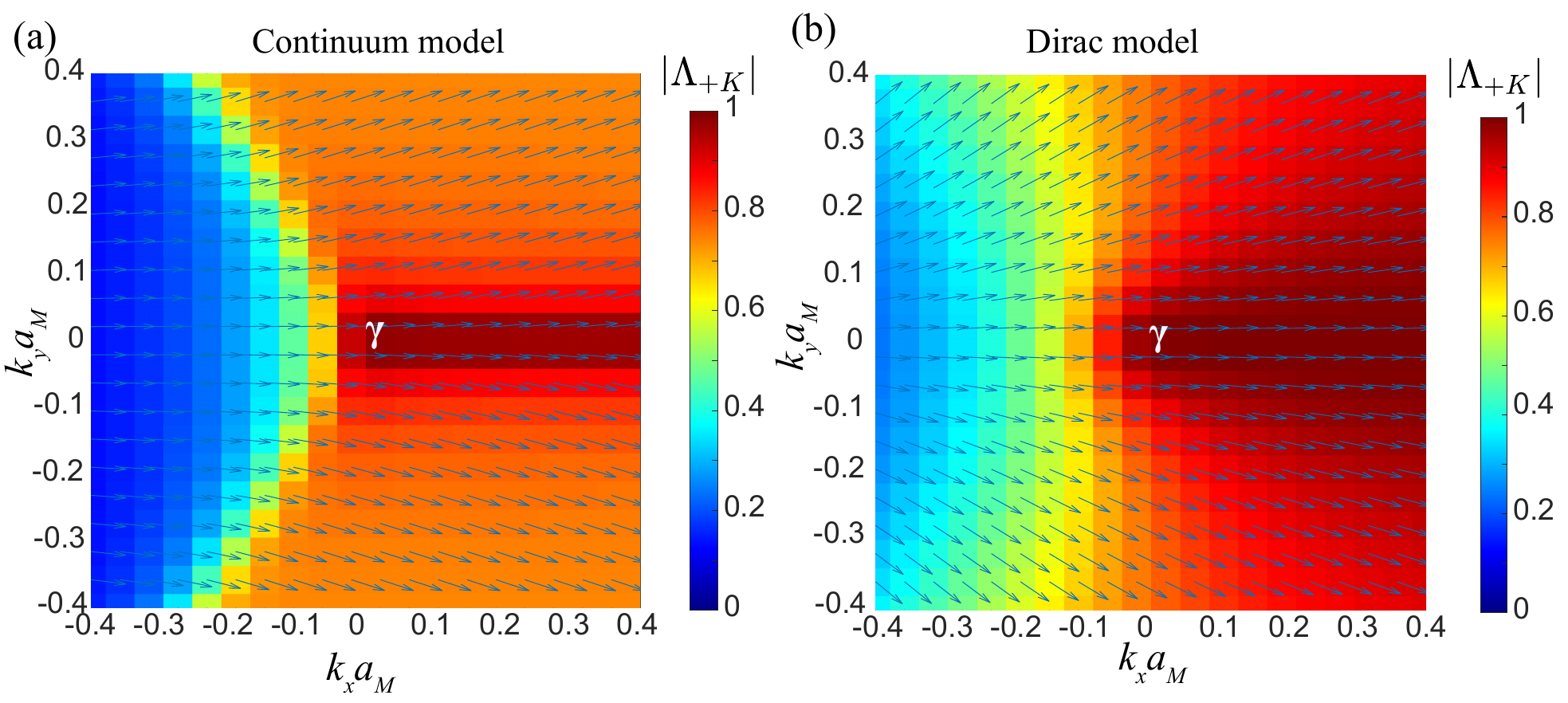}
\caption{(a) and (b) show the form factor calculated from the continuum and Dirac models respectively. The color bar represents the modulus of the form factor at the K valley, and the arrow indicates the phase. The moir\'e potential used in the continuum model is $V_{b}=1$ meV, and the massive gap in the Dirac model is $\Delta_{M}=1$ meV.}
\label{FIG_S1}
\end{figure}

\section{ The numerical  Hartree-Fock calculation for the minimal model with massive Dirac} To verify the understanding from the above analytical mean-field analysis, we have constructed an effective three-band model and performed a numerical Hartree-Fock calculation in the main text. Here, we present the corresponding details here. This three-band model in the basis ($\ket{\gamma,1+},\ket{\gamma,2+},\ket{\kappa,3-}$) reads

\begin{align}
&H_{MF}(\bm{k})=H^{eff}_0(\bm{k})+\Delta(\bm{k}),\nonumber\\
&H^{eff}_0(\bm{k})=\begin{pmatrix}
	H_{+,\gamma}(\bm{k})-\mu&0\\
	0&\xi_{-}(\bm{k})
\end{pmatrix},\\
&\Delta(\bm{k})=\begin{pmatrix}
	\Delta_{11}(\bm{k})&\Delta_{12}(\bm{k})&\Delta_{13}(\bm{k})\\
	\Delta_{21}(\bm{k})&\Delta_{22}(\bm{k})&\Delta_{23}(\bm{k})\\
	\Delta_{31}(\bm{k})&\Delta_{32}(\bm{k})&\Delta_{33}(\bm{k})
\end{pmatrix}.
\end{align}

As  the Hartree  energy is negligible in our study, we define the order parameters $\Delta_{ij}$ in terms of the Fock energy as

\begin{equation}
\Delta_{ij}(\bm{k})=-\frac{1}{A}\sum_{\bm{k'}}\tilde{V}_{ij}(\bm{k}-\bm{k'})(\rho_{ij}(\bm{k})-\rho_0(\bm{k})).\label{Eq_sf}
\end{equation}
Here,    $\tilde{V}_{ij}(\bm{k}-\bm{k'})$ denotes the effective interaction between the electrons.  For simplicity, we set the interaction  $\tilde{V}_{ij}(\bm{k}-\bm{k'})=V_I/\sqrt{(\bm{k}-\bm{k'})^2+\lambda^2}$. $\rho_{ij}(\bm{k})=\braket{c_{i}^{\dagger}(\bm{k})c_j(\bm{k})}$ is the the density matrix,  $\rho_0(\bm{k})$ is the density matrix in the noninteracting limit with $\mu$  at half-filling. The density matrix  in each iteration 
is calculated by diagonalizing $H_{MF}(\bm{k})$ and obtaining $H_{MF}(\bm{k})\ket{\psi_n(\bm{k})}=E_n(\bm{k})\ket{\psi_n(\bm{k})}$, where  $\ket{\psi_n(\bm{k})}=\sum_j U_{jn}(\bm{k})\ket{c_j(\bm{k})}$. The density matrix in this quasi-particle band basis thus is evaluated as 
\begin{equation}
\rho_{ij}(\bm{k})=\sum_n U^*_{j,n}(\bm{k})U_{i,n}(\bm{k})f(E_n(\bm{k})).\label{Eq_rho}
\end{equation}
Here, $f$ is the Fermi distribution function at zero temperature.    Then we can perform a Hartree-Fock mean-field self-consistent calculation to obtain $\Delta(\bm{k})$  with different interaction strengths according to Eq.~(\ref{Eq_sf}) and Eq.~(\ref{Eq_rho}).

\section{The effective interaction from the form factor}

The effective interaction $\tilde{V}^{+-}_{\bm{k',\bm{k}+\bm{Q},\bm{k}-\bm{k'}}}$ is related to the form factor of $\tau=+$ valley $\Lambda_{\bm{G}}^{+}(\bm{k},\bm{k'})$ and  $\tau=-$ valley $\Lambda_{\bm{G'}}^{-}(\bm{k'}+\bm{Q},\bm{k}+\bm{Q})$. It is straightforward to evaluate $\Lambda_{\bm{G'}}^{-}(\bm{k'}+\bm{Q},\bm{k}+\bm{Q})$.   As we have pointed out, it is the maximum of the top moir\'e band at the $\tau=-$ valley locating near Fermi energy, where the wave function  $\ket{\bm{k}+\bm{Q},-}=\sum_{\bm{G}}U_{\bm{k}+\bm{Q}}(\bm{G})e^{i(\bm{k}+\bm{Q}+\bm{G})\cdot{\bm{r}}}\delta_{\bm{G},0}=e^{i(\bm{k}+\bm{Q})\cdot\bm{r}}$. As a result, the form factor of the $-$ valley can be obtained as
\begin{equation}
\Lambda_{\bm{G'}}^{-}(\bm{k'}+\bm{Q},\bm{k}+\bm{Q})=\delta_{\bm{G'},0}.\label{Form1}
\end{equation} 
Note that we have taken the top band moir\'e band so that $n=1$.  


Next, we  calculate the form factor $\Lambda_{\bm{G}}^{+}(\bm{k},\bm{k'})$ with the two-band effective Hamiltonian  we obtained.  According to the massive Dirac Hamiltonian Eq.~\eqref{massive_Dirac},  the eigenenergy representing the top moir\'e band  is $E_{+}(\bm{k})=\epsilon_{0}+\sqrt{v_F^2k^2+\Delta^2}$, and
the corresponding eigenstates are
\begin{eqnarray}
&&\ket{E_{+}(\bm{k})}= (\cos \frac{\theta_{k}}{2} \ket{\epsilon_{1}}-\sin \frac{\theta_{k}}{2}\CP{e^{-i\varphi_{\bm{k}}}}\ket{\epsilon_{3}})e^{i\bm{k}\cdot\bm{r}}\nonumber\\&&=\sum_{j=0,1,2}U_{\bm{k}+\bm{\gamma_3}}(\bm{G}_{j})e^{i(\bm{k}+\bm{\gamma_3}+\bm{G}_j)\cdot \bm{r}},
\end{eqnarray}
Here, $\bm{k}$ is measured with respect to $\bm{\gamma}$ point, $\theta_{k}=\tan^{-1}\frac{v_F k}{\Delta_M}$ with $k=\sqrt{k_x^2+k_y^2}$, and $\varphi_{\bm{k}}=\tan^{-1}\frac{k_y}{k_x}$, the moir\'e reciprocal lattice vectors $\bm{G_0}=(0,0), \bm{G}_1=(-\frac{3}{2},\frac{\sqrt{3}}{2})\kappa$, $\bm{G}_2=(-\frac{3}{2},-\frac{\sqrt{3}}{2})\kappa$  and the coefficients 
\begin{equation}
U_{\bm{k}+\bm{\gamma_3}}(\bm{G}_j)=\frac{1}{\sqrt{3}}(\cos\frac{\theta_{k}}{2}-\omega^{-j}\sin\frac{\theta_{k}}{2}\CP{e^{-i\varphi_{\bm{k}}}}).
\end{equation}
When the Coulomb interaction $V(\bm{q})$ is dominant by the long wave limit ($\bm{q}$ is close to zero), we can keep only $\Lambda_{\bm{G=0}}^{+}$ for our discussion. 
Then we find the nonvanishing form factors related to the states near $\gamma$ point are
\begin{eqnarray}
&&\Lambda^{+}_{\bm{G}=0}(\bm{k},\bm{k'})=\sum_{j=0,1,2}U^*_{\bm{k}+\bm{\gamma_3}}(\bm{G}_j)U_{\bm{k'}+\bm{\gamma_3}}(\bm{G}_j)\nonumber\\&&=\cos\frac{\theta_{k}}{2}\cos\frac{\theta_{k'}}{2}+\sin\frac{\theta_{k}}{2}\sin\frac{\theta_{k'}}{2}e^{i(\varphi_{\bm{k}}-\varphi_{\bm{k'}})}\label{Form2}
\end{eqnarray}

Then substituting Eq.~(\ref{Form1}) and Eq.~(\ref{Form2}) into Eq.~\eqref{eff_interaction}, we find the effective interaction to the leading order is given by
\begin{eqnarray}
&&V_{eff}(\bm{k}-\bm{k'})=V_0(\bm{k}-\bm{k'})(\cos\frac{\theta_{k}}{2}\cos\frac{\theta_{k'}}{2}+\nonumber\\&&\sin\frac{\theta_{k}}{2}\sin\frac{\theta_{k'}}{2}\YM{e^{i(\varphi_{\bm{k}}-\varphi_{\bm{k'}})}}).
\end{eqnarray}
Here, we define the interaction strength $V_0(\bm{k}-\bm{k'})=\frac{2\pi e^2}{\epsilon\sqrt{|\bm{k}-\bm{k'}|^2+\lambda^2}}$.
The self-consistent Eq.~\eqref{Self_equation} thus becomes

\begin{eqnarray}
\Delta^{IVC}_{+-}(\bm{k})&&=\cos\frac{\theta_{k}}{2}\sum_{\bm{k'}}V_0(\bm{k}-\bm{k'})F_1(\bm{k'})\nonumber\\
&&+\sin\frac{\theta_{k}}{2}\YM{e^{i\varphi_{\bm{k}}}}\sum_{\bm{k'}}V_0(\bm{k}-\bm{k'})F_2(\bm{k'})  
\end{eqnarray}
where we define
\begin{eqnarray}
F_1(\bm{k'})&&=\frac{\cos\frac{\theta_{k'}}{2}\Delta_{+-}^{IVC}(\bm{k'})}{2\sqrt{\xi'(\bm{k'})^2/4+|\Delta_{+-}^{IVC}(\bm{k'})|^2}},\\ F_2(\bm{k'})&&=\frac{\sin\frac{\theta_{k'}}{2}\YM{e^{-i\varphi_{\bm{k'}}}}\Delta_{+-}^{IVC}(\bm{k'})}{2\sqrt{\xi'(\bm{k'})^2/4+|\Delta_{+-}^{IVC}(\bm{k'})|^2}}.
\end{eqnarray}
with $\xi'(\bm{k'})=(\tilde{\xi}_{+}(\bm{k'})-\tilde{\xi}_{-}(\bm{k'}+\bm{Q}))/2$.  Notably, \YM{the first term represents $s$-wave channel while the \YM{second} term gives rise to the $p+ip$-wave channel. We assume the  $C_3$ symmetry is not spontaneously broken so that we can decouple the self-consistent equation into  $s$ and $p$ channel:  }
\YM{
\begin{align}
	\Delta_s^{IVC}(\bm{k})&=\cos \frac{\theta_{\bm{k}}}{2}\sum_{\bm{k'}} V_0(\bm{k}-\bm{k'}) F_1(\bm{k'}),\\
	\Delta_{p+ip}^{IVC}(\bm{k})=&\sin \frac{\theta_{\bm{k}}}{2}\sum_{\bm{k'}} V_0(\bm{k}-\bm{k'}) F_2(\bm{k'}).
	\end{align}}
	where the $\Delta_{+-}^{IVC}(\bm{k'})$ in $F_1(\bm{k'})$ and $F_2(\bm{k'})$  is replaced as $\Delta_{s}^{IVC}(\bm{k'})$ and $\Delta_{p+ip}^{IVC}(\bm{k'})$, respectively. We can consider  $(\bm{k}-\bm{k'})\ll \lambda$ so that $V_0(\bm{k}-\bm{k'})\approx V_0=\frac{2\pi e^2}{\epsilon\lambda}$. In this case,
	\begin{align}
\Delta^{IVC}_{s}(\bm{k})=\Delta_s\cos\frac{\theta_{\bm{k}}}{2},\\  \Delta^{IVC}_{p}(\bm{k})=\Delta_p\sin\frac{\theta_{\bm{k}}}{2}e^{i\varphi_{\bm{k}}}.
\end{align}  
Then we can obtain:
\begin{eqnarray}
&&V_0\sum_{\bm{k'}}\frac{\sin^2\frac{\theta_{\bm{k'}}}{2}}{{2\sqrt{\xi'(\bm{k'})^2/4+\Delta_s^2\sin^2\frac{\theta_{\bm{k'}}}{2}}}}=1,\\  &&V_0\sum_{\bm{k'}}\frac{\cos^2\frac{\theta_{\bm{k'}}}{2}}{{2\sqrt{\xi'(\bm{k'})^2/4+\Delta_p^2\cos^2\frac{\theta_{\bm{k'}}}{2}}}}=1.
\end{eqnarray}

\YM{In the BCS limit where the interaction strength is small comparing with the band width $\Lambda_{c}$ , we can replace $\sum_{\bm{k'}}$ as $\int d\xi' N_0$,  and taking the integration with respect to $\xi'$, where $N_0$ is the density of states near the Fermi energy. In the limit of $v_Fk_f\gg \Delta_M$,  $\theta_{k}\rightarrow \pi/2$, we find  
\begin{equation}
	\Delta_s=\Delta_p=2\Lambda_ce^{-\frac{1}{N_0\tilde{V_0}}}.
	\end{equation}}

	
	
	\section{Charge density wave behavior induced by the finite momentum Q} As we have pointed out in the main text, the order parameter $\braket{c^{\dagger}_{+,\bm{k}}c_{-,\bm{k}+\bm{Q}}}$ would display as a charge density wave behavior in real space, where $\bm{Q}=(\frac{4\pi}{3a_M},0)$ is a momentum connecting the center and corner of the moir\'e Brillouin zone.  This finite $\bm{Q}$ breaks the original translational symmetry. As a result,  the  Brillouin zones are folded into one-third of the original one (see Fig.~\ref{fig:fig4}a), while the unit cell is enlarged to three times the original moir\'e unit cell. 
	
	To show this charge density wave behavior explicitly, we calculate the local density state by adopting finite-$\bm{Q}$ IVC order parameter with near half-filling.  Being consistent with the main text, we only added the IVC order parameter onto the top two  moir\'e bands to generate an IVC gap near half-filling, which is written as  $\Delta_{IVC}(\bm{k})\ket{\psi_{n=1,+}(\bm{k})}\bra{\psi_{n=1,-}(\bm{k}+\bm{Q})}+\Delta^*_{IVC}(\bm{k})\ket{\psi_{n=1,-}(\bm{k}+\bm{Q})}\bra{\psi_{n=1,+}(\bm{k})}$. Here $\ket{\psi_{n\tau}(\bm{k})}$ is obtained by digonalizing $H_{\tau}(\bm{r})$ in plane wave basis ($n=1$ label the top moir\'e band), and the IVC order parameter $\Delta_{IVC}(\bm{k})=\Delta_{IVC}(k_x+ik_y$).  Then we can diagonalize the new mean-field Hamiltonian that includes the IVC order parameter between two valleys, and obtain the eigenenergy $E_{n\bm{k}}$ and the eigenstate   $\psi_{n\bm{k},\bm{Q}}(\bm{r})=\sum_{\bm{G}}U_{n\bm{k},\tau=+}(\bm{G})e^{i(\bm{k}+\bm{G})\cdot\bm{r}}+\sum_{\bm{G}}U_{n\bm{k},\tau=-}(\bm{G})e^{i(\bm{k}+\bm{Q}+\bm{G})\cdot \bm{r}}$. 
	The local density of states at  energy $E$ and  position $\bm{r}$ is given by
	\begin{equation}
\rho(E,\bm{r})=\frac{1}{A}\sum_{n,\bm{k}}|\psi_{n\bm{k},\bm{Q}}(\bm{r})|^2\delta (E-E_{n\bm{k}}).
\end{equation}

The calculated local density of states near half-filling in the case without IVC state ($\Delta_{IVC}=0$)  and with IVC state ($\Delta_{IVC}=1$ meV) are plotted in Figs.~\ref{fig:fig4}b and~\ref{fig:fig4}c, respectively. It can be clearly seen that the presence of the finite $\bm{Q}$ generates a charge density wave behavior, which enlarges the unit cell to be $\sqrt{3} a_M\times \sqrt{3} a_M$ in real space. Note that the pattern of the local density of states within each unit cell depends on the detailed parameters of the moir\'e potential, but the periodicity of the density wave would not be affected and can be easily verified experimentally through probing the local density of state with the STM.
\end{appendix}


%

\end{document}